\documentclass[10pt,twocolumn,english,aps,manuscript]{revtex4}
\usepackage{mathptmx}
\usepackage[T1]{fontenc}
\usepackage[latin9]{inputenc}
\usepackage{amsbsy}
\usepackage{amstext}
\usepackage{graphicx}

\makeatletter

\newcommand{\noun}[1]{\textsc{#1}}
\providecommand{\tabularnewline}{\\}

\@ifundefined{textcolor}{}
{%
 \definecolor{BLACK}{gray}{0}
 \definecolor{WHITE}{gray}{1}
 \definecolor{RED}{rgb}{1,0,0}
 \definecolor{GREEN}{rgb}{0,1,0}
 \definecolor{BLUE}{rgb}{0,0,1}
 \definecolor{CYAN}{cmyk}{1,0,0,0}
 \definecolor{MAGENTA}{cmyk}{0,1,0,0}
 \definecolor{YELLOW}{cmyk}{0,0,1,0}
}

\makeatother

\usepackage{babel}
\begin{document}

\title{Coherent exciton dynamics in supramolecular light-harvesting nanotubes
revealed by ultrafast quantum process tomography}

\author{Joel Yuen-Zhou$^{1}$, Dylan H. Arias$^{1,2}$, Dorthe M. Eisele$^{1,2}$,
Colby P. Steiner$^{1,2}$, Jacob J. Krich$^{3}$, Moungi Bawendi$^{1,2}$,
Keith A. Nelson$^{1,2}$ and Alán Aspuru-Guzik$^{1,4}$}

\affiliation{$^{1}$Center for Excitonics, Research Laboratory of Electronics,
Massachusetts Institute of Technology, Cambridge, MA, USA,}

\affiliation{$^{2}$Department of Chemistry, Massachusetts Institute of Technology,
Cambridge, MA, USA,}

\affiliation{$^{3}$Department of Physics, University of Ottawa, Ottawa, ON, Canada,}

\affiliation{$^{4}$Department of Chemistry and Chemical Biology, Harvard University,
Cambridge, MA, USA.}
\begin{abstract}
Long-lived exciton coherences have been recently observed in photosynthetic
complexes via ultrafast spectroscopy, opening exciting possibilities
for the study and design of coherent exciton transport. Yet, ambiguity
in the spectroscopic signals has led to arguments for interpreting
them in terms of the exciton dynamics, demanding more stringent tests.
We propose a novel strategy, Quantum Process Tomography (QPT) for
ultrafast spectroscopy, to reconstruct the evolving quantum state
of excitons in double-walled supramolecular light-harvesting nanotubes
at room temperature. The protocol calls for eight transient grating
experiments with varied pulse spectra. Our analysis reveals unidirectional
energy transfer from the outer to the inner wall excitons, absence
of nonsecular processes, and an unexpected coherence between those
two states lasting about 150 femtoseconds, indicating weak electronic
coupling between the walls. Our work constitutes the first experimental
QPT in a ``warm'' and complex system, and provides an elegant scheme
to maximize information from ultrafast spectroscopy experiments.
\end{abstract}
\maketitle
Recently, there has been great excitement about the detection of long-lived
coherent dynamics in natural light-harvesting photosynthetic complexes
via two-dimensional spectroscopy \cite{engelfleming,scholes,engelchicago}.
This long-lived coherence has generated interest and debate about
its role in the efficient design of light-harvesting and exciton transport
in biological and artificial settings \cite{mohseni,plenio,BrumerShapiroPNAS,kassal}.
These discussions have highlighted the importance of correctly interpreting
the spectroscopic signals in terms of the microscopic dynamics in
the material. The interplay between excitonic dynamics and vibrational
dynamics can produce complex and potentially ambiguous spectroscopic
signals, which can make extraction of information about exciton transport
challenging \cite{cinabiggs1,witness,hauer}. Therefore, it is essential
to develop methods to reliably extract the quantum dynamics of the
interrogated material. In this article, we demonstrate the systematic
characterization of the quantum dynamics of a condensed phase molecular
system, namely, the excitons originating from the inner and outer
walls of supramolecular light-harvesting nanotubes, via ultrafast
Quantum Process Tomography (QPT) \cite{procedia,yuen-aspuru,yuenzhou}.
This manuscript is organized as follows: First, we briefly sketch
the QPT formalism as a general method to maximize information from
a quantum system interacting with its environment. Then, we describe
the nanotubes and the optical setup, and explain how these two are
ideally suited for the QPT protocol. Finally, we present the experimental
data and its analysis, yielding a full characterization of the quantum
dynamics of the excitonic system. To our knowledge, this article constitutes
the first experimental realization of QPT on a molecular system in
condensed phase, and provides general guidelines to adapt standard
spectroscopic experiments to carry out QPT.

The time evolution of the excited state of an open quantum system
(a system interacting with its environment, e.g., an electronic system
interacting with an environment of vibrations) that is prepared by
a pump pulse is, under general assumptions, given by \cite{nielsenchuangbook,yuen-aspuru,yuenzhou},

\begin{equation}
\rho(T)=\chi(T)\rho(0),\label{eq:linear_transformation-1}
\end{equation}
where $\rho(T)$ is the density matrix of the system at time $T$
after the pump pulse, and the\noun{ }\emph{process matrix} $\chi(T)$
is a propagator that relates input and output states. By introducing
a basis, Eq. (\ref{eq:linear_transformation-1}) reads $\rho_{qp}(T)=\sum_{ij}\chi_{qpij}(T)\rho_{ij}(0)$,
where $\chi_{qpij}(T)$ denotes a transition probability amplitude
of ending in state $|q\rangle\langle p|$ at time $T$ having started
in state $|i\rangle\langle j|$. In other words, $\chi(T)$ characterizes
the transfer processes amongst populations (diagonal elements of $\rho$)
and coherences (off-diagonal elements of $\rho$). This phenomenology
is familiar in nonlinear spectroscopy and can be discussed in terms
of Double-Sided Feynman diagrams \cite{mukamel,mukamel_2d,minhaengbook}.
The process matrix $\chi(T)$ is a linear transformation of $\rho(0)$,
which in turn yields the remarkable observation that, once $\chi(T)$
is given, the dynamics of the system are completely characterized;
they are valid for arbitrary system initial states, including any
interaction with the environment, whether characterized by Markovian
or non-Markovian processes. The reconstruction of $\chi(T)$ is the
central goal of QPT, an essential step in the verification of quantum
technologies \cite{blatttraps,childs,cory,natphysqpt,opticallattices,solidstate,rabitz,cinaprl,avisar}
and dynamical models. Determining $\chi(T)$ ensures that we have
extracted the maximal amount of information possible about the excited
state system dynamics. Previous theoretical work showed that selectively
preparing and measuring a number of linearly independent initial states
via laser excitation suffices to accomplish QPT \cite{yuen-aspuru,yuenzhou,hoyer-whaley,rabitz_tomography}.
Hence, QPT can in principle be realized with the tools of ultrafast
spectroscopy by collecting sufficient number of signals with varying
frequency, polarization, and time delays. This work represents the
first realization of QPT in a ``warm'' and complex system.

\section*{Results}

We study the exciton states of light-harvesting nanotubes (Fig. \ref{fig:system}a
and SI Sec. 1) that self-assemble in a water/methanol solution from
the amphiphilic cyanine dye monomer 3,3'-bis(2-sulfopropyl)-5,5',6,6'-tetrachloro-1,1'-dioctylbenzimidacarbocyanine
\cite{derossi} (abbreviated as C8S3) at room temperature (298 K).
The nanotubes are about 10 nm in diameter, several micrometers long,
and have a remarkably uniform supramolecular structure \cite{eiselenanotech}:
they are composed of two concentric cylinders\textemdash{}an inner
wall cylinder and an outer wall cylinder\textemdash{}separated by
about 4 nm \cite{vonberlepsch,eiselejacs}.

Upon self-assembly, the broad absorption band of the monomer (Fig.
\ref{fig:system}b) undergoes a large redshift of $\text{\textasciitilde}$2,500
cm$^{-1}$ reflecting the strong coupling of the molecular transition
dipole moments forming delocalized excitonic eigenstates \cite{scholesnatchem}.
In addition, a complex pattern of absorption bands occurs, caused
by the nanotube\textquoteright{}s complex cylindrical geometry \cite{DidragaKlugkistJPCB2002,didraga}.
Bands (1) at $\sim16600\,\mbox{cm}^{-1}$ and (2) at $\sim17100\,\mbox{cm}^{-1}$
are polarized primarily parallel to the cylindrical axis and correspond
to transitions which couple the Ground State Manifold (GSM, $|g\rangle$,
state with no excitations) and the Single Exciton Manifold (SEM),
composed of $|I\rangle$ and $|O\rangle$, that is, states that concentrate
exciton amplitude mostly on the inner wall and the outer wall cylinders,
respectively \cite{eisele}. These transitions occur at $\omega_{Ig}\sim16600$
cm$^{-1}$ and $\omega_{Og}\sim17100$ cm$^{-1}$ ($\omega_{ij}=\omega_{i}-\omega_{j}$
denotes a difference in energies). The rest of the bands (shoulder
at higher energies than band (2), not labeled in Fig. \ref{fig:system}b)
are polarized along the equatorial plane of the nanotubes. By flowing
the nanotubes through a cell, they align their long axes with the
direction of the flow. Therefore, polarized light parallel to the
flow can be used to isolate the transitions to $|I\rangle$ and $|O\rangle$,
yielding the simplified absorption spectrum in Fig. \ref{fig:lin_spectra}a.

\begin{center}
\begin{figure}
\begin{centering}
\includegraphics[scale=0.4]{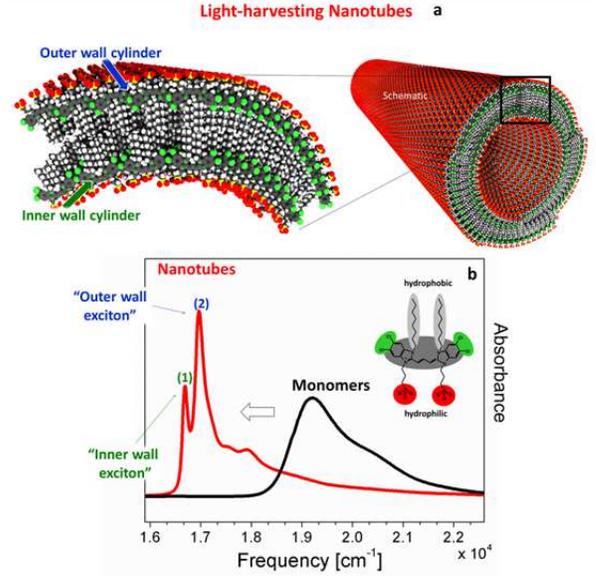}
\par\end{centering}

\caption{The excitonic system under consideration: Light-harvesting nanotube
consisting of a double-walled cylindrical aggregate of amphiphilic
cyanine dye molecules. (a) Schematic of the self-assembled light-harvesting
nanotube (for clarity using only one molecule per unit cell): double-walled
morphology with the hydrophilic sulfonate groups (red) on the exterior,
the hydrophobic alkyl chains (light grey) in the interior of the bilayer
and the cyanine dye chromophore (dark grey). (b) Absorption spectra
of amphiphilic dye monomers C8S3 (black) dissolved in methanol (no
aggregation) and nanotubular prepared in water/methanol (red). The
nanotube\textquoteright{}s inner-wall and outer-wall cylinders featuring
distinct delocalized exciton bands (1) and band (2) associated with
the $|I\rangle$ and $|O\rangle$ excitons. (Reprinted with permission
from Eisele, D.M., et al., Nat. Nanotech. (4): 658-663, 2009 and Nat.
Chem. (4): 655-662, 2012. Copyright Nature Publishing Group).\label{fig:system}}
\end{figure}

\par\end{center}

The well separated peaks of $|I\rangle$ and $|O\rangle$ (Fig. \ref{fig:system}b)
suggest a QPT scheme where selectivity is achievable by varying the
carrier frequencies of the pulses and fixing their polarizations to
be along the long axes of the nanotubes. In particular, we work within
a transient grating (TG) setup, where three weak intensity non-collinear
narrowband beams with wavectors $\boldsymbol{k}_{1}$, $\boldsymbol{k}_{2}$,
and $\boldsymbol{k}_{3}$ interact with the nanotubes, and the coherent
signal diffracted at $\boldsymbol{k}_{s}=-\boldsymbol{k}_{1}+\boldsymbol{k}_{2}+\boldsymbol{k}_{3}$
is spectrally interfered with a broadband local oscillator (LO) fourth
pulse at $\boldsymbol{k}_{4}=\boldsymbol{k}_{s}$, generating a complex
(absorptive and dispersive) spectrum as a function of waiting time
$T=t_{3}-t_{2}$ ($t_{i}$ denotes the arrival time of each pulse)
(SI Sec. 2). Pump pulses 1 and 2 interact simultaneously $(t_{1}=t_{2})$
with the sample. The first three narrowband pulses are chosen from
a toolbox of two different pulse shapes, namely, a pulse that exclusively
excites $|I\rangle$ and another one that excites $|O\rangle$, which
we shall label as I and O, respectively. This generates eight different
experiments associated with the triads of carrier frequencies: OOO,
OOI, III, IIO, OIO, OII, IOI, and IOO. Fig. \ref{fig:lin_spectra}a
shows the spectra of the pulses on top of a magnified version of the
absorption spectrum of the material from Fig. \ref{fig:system}c.

We are interested to probe the dynamics of the SEM. In the TG experiment,
the first two pulses prepare an initial SEM state, which then evolves
for a waiting time $T$ \cite{yuenzhou}. The third pulse probes the
state at time $T$ by inducing Stimulated Emission (SE) from the SEM
to the GSM or Excited State Absorption (ESA) to the Doubly Excited
Manifold (DEM). The DEM consists of three states with two excitons
each: $|II\rangle$, $|IO\rangle$, and $|OO\rangle$, whose energies
we assign as being the sums of the corresponding single-exciton states,
with no binding energies, this being a reasonable assumption for molecular
excitons (see SI Sec. 4). We also detect the reduced absorption of
the third pulse from the ground state $|g\rangle$ (due to the population
moved to the SEM), known as Ground State Bleach (GSB). Finally, the
decay of this bleach is Ground State Recovery (GSR), which contributes
as the population in the SEM decays back to the GSM.

Fig. \ref{fig:lin_spectra}b shows the energy-level diagram for our
system, as determined self-consistently from the TG spectra (see SI
Sec. 4). The rationale of our QPT scheme is the following (illustrated
in Figs. 2c and 3): Narrowband optical pulses allow us to selectively
create populations or coherences in the SEM. For instance if $(\omega_{1}=\omega_{2})=(\omega_{Og},\omega_{Ig})$,
the initial state at the beginning of the waiting time will be $\rho(0)=|I\rangle\langle O|$
(in the rotating wave approximation (RWA), pulse 1 ``acts on the
bra'' and pulse 2 ``acts on the ket'' \cite{mukamel,minhaengbook}).
This state evolves for the waiting time $T$, when the third pulse
and the diffracted probe light can detect it.

\begin{center}
\begin{figure*}
\begin{centering}
\includegraphics[scale=0.33]{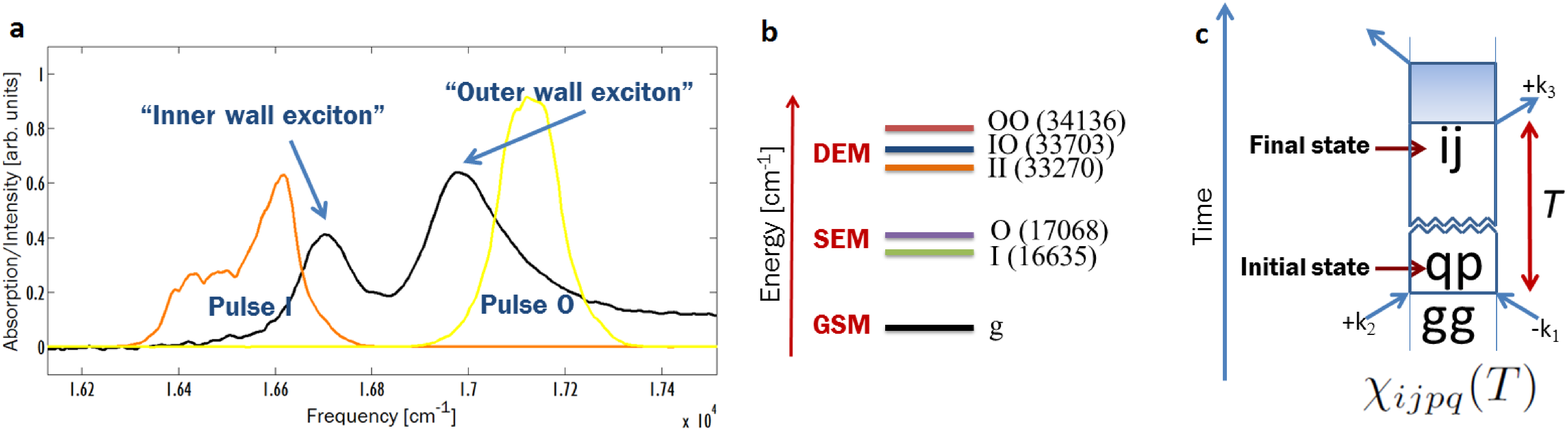}
\par\end{centering}

\caption{The concepts behind our QPT protocol. (a) Simplified absorption spectrum
of the light harvesting nanotubes in the flow cell revealing only
two optical transitions when exposed to light that is polarized along
the long axes of the nanotubes. Each of the three pulses in each TG
experiment is narrowband enough that it is selective towards the $\{|g\rangle\leftrightarrow|I\rangle,|I\rangle\leftrightarrow|II\rangle,|O\rangle\leftrightarrow|IO\rangle\}$
or the $\{|g\rangle\leftrightarrow|O\rangle,|O\rangle\leftrightarrow|OO\rangle,|I\rangle\to|IO\rangle\}$
transitions, respectively. (b) Energy level diagram of the system.
Transitions are allowed between the Ground-State Manifold and any
state in the Singly-Excited Manifold, or between any state in the
latter and any in the Doubly-Excited Manifold. (c) Double-sided Feynman
Diagram representing the general idea of the QPT protocol using TG
experiments. The first two pulses prepare the initial state and the
last two pulses detect the final state at the end of the waiting time
$T$. \label{fig:lin_spectra}}
\end{figure*}

\par\end{center}

Fig. \ref{fig:protocol} exhaustively enumerates the possible initial
states prepared by pulses 1 and 2 and the possible final states detected
by pulses 3 and 4, and hence, lists the elements of $\chi(T)$ that
are measured by keeping track of each peak in the series of frequency-resolved
TG spectra as a function of $T$. The emission frequencies are associated
with the final elements in each Feynman diagram. For instance, let
us consider the experiment OIO. Pulses 1 and 2 selectively prepare
$|I\rangle\langle O|$, and this state evolves for a time $T$. There
could potentially be nonzero probability amplitudes $\chi_{OOIO}(T)$
and $\chi_{IIIO}(T)$ of population being transferred into $|I\rangle\langle I|$
or $|O\rangle\langle O|$. These processes can be detected with the
third pulse O, inducing the SE transition $|O\rangle\langle O|\to|O\rangle\langle g|$
and the ESA transitions $|O\rangle\langle O|\to|OO\rangle\langle O|$,
$|I\rangle\langle I|\to|IO\rangle\langle I|$, all of which emit at
$\omega_{Og}=\omega_{OO,O}=\omega_{IO,I}=17068\,\mbox{cm}^{-1}$ in
the corresponding TG spectrum. Similarly, these same elements of $\chi(T)$
contribute to the peak at $\omega_{Ig}=\omega_{II,I}=\omega_{IO,O}=16635\,\mbox{cm}^{-1}$
of the experiment OII.

\begin{center}
\begin{figure*}
\begin{centering}
\includegraphics[scale=0.26]{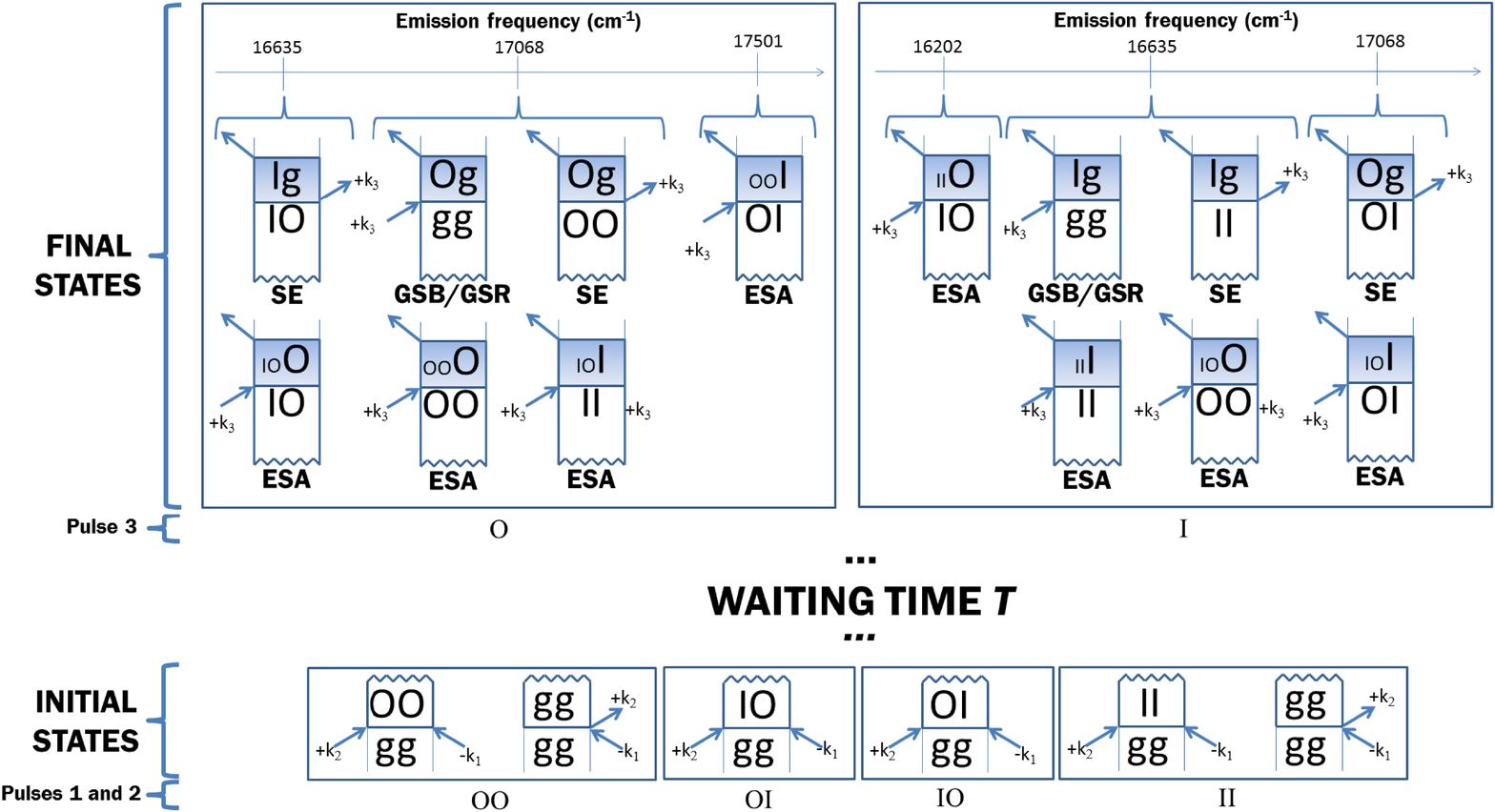}
\par\end{centering}

\caption{QPT protocol for the two-band exciton system of the double-walled
J-aggregate. In the TG setup, the carrier-frequencies of the first
two narrowband pulses (bottom) selectively determine the possible
initial states. Due to interactions with the vibrational surroundings
(the bath), the initial state of the excitons can potentially transfer
into other states of the SEM during the waiting time $T$. Narrowband
pulse 3 and broadband LO pulse 4 detect these transfers by producing
a frequency-resolved TG spectrum which features a set of emission
frequencies that correlate with the state of the system at the end
of the waiting time, just as depicted in this figure. \label{fig:protocol}}
\end{figure*}

\par\end{center}

Fig. \ref{fig:raw_data} shows the data obtained from the eight frequency-resolved
TG experiments as a function of waiting time $T$. The data for $T>500\,\mbox{fs}$
were not included in the analysis due to the increasing influence
of pulse intensity roll-off as a function of delay in our pulse shaping
apparatus \cite{colbert}. The below analysis indicates that the coherent
dynamics are complete by 500 fs (see Fig. \ref{fig:tomo}). Both absorptive
and dispersive (in our phase convention, real and imaginary, respectively)
parts of the complex valued spectra are collected, but we only show
the real part. Whereas Fig. \ref{fig:protocol} predicts that three
peaks in frequency domain are possible in each of the spectra, we
find surprisingly that there is only one peak of significant amplitude
in each spectrum, revealing that nonsecular processes such as coherence
to population transfers are negligible or too small to be detected
with the current experimental setup. Yet, as noted in the previous
paragraph as well as in \cite{yuen-aspuru,yuenzhou,witness} and SI
Sec. 3, some of the peaks report on more than one element of $\chi(T)$,
and a more careful procedure to dissect their contributions is necessary.
In fact, each peak amplitude can be expressed as a linear combination
of elements of $\chi(T)$ where the coefficients are products of transition
dipole moments. We extract the required information about the dipoles
self-consistently from the TG data via the initial condition $\chi_{ijqp}(0)=\delta_{iq}\delta_{jp}$
(see SI, Sec. 5). The information associated with $\chi(T)$ is then
obtained by integrating the area under the complex valued peaks and
carrying out a constrained linear inversion procedure. This procedure
is a semidefinite programming routine \cite{gb08,cvx} that ensures
that the extracted $\chi(T)$ maps physical density matrices as inputs
(Hermitian, trace preserving, and positive) to physical density matrices
as outputs (SI Sec. 5).

\begin{center}
\begin{figure*}
\begin{centering}
\includegraphics[scale=0.36]{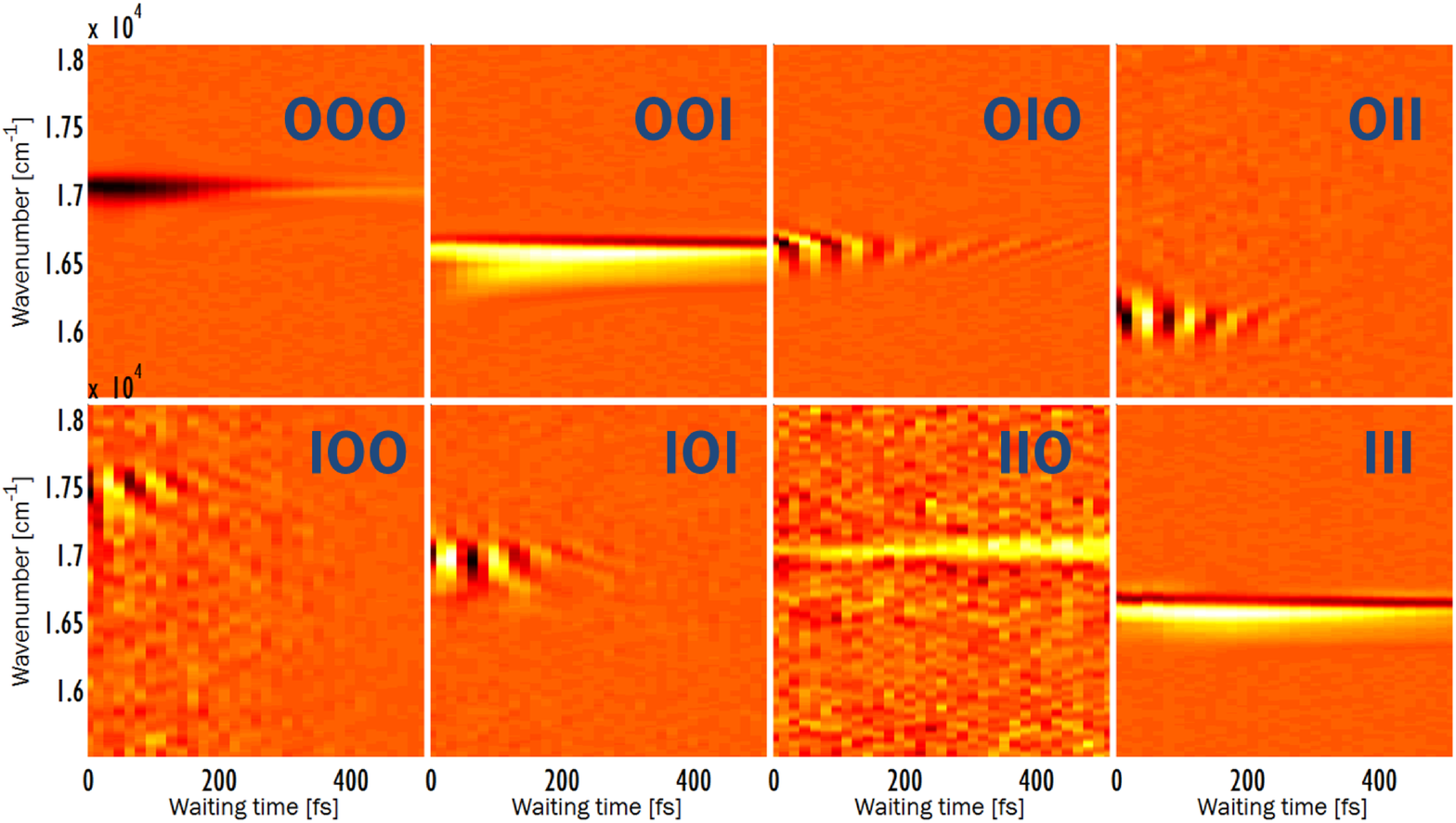}
\par\end{centering}

\caption{Absorptive part of eight narrowband TG experiments on the two exciton-band
system of the double-wall J-aggregate. The data only show one significant
peak per spectrum (instead of a maximum of three, as outlined in Fig.
\ref{fig:protocol}). Population transfer is revealed in the OOO,
OOI, IIO, and III panels, whereas coherence dynamics are monitored
by OIO, OII, IOO, and IOI. Coherence between $|I\rangle$ and $|O\rangle$
lasts for about 150 fs at room temperature and observed as fringes
as a function of waiting time $T$. The dispersive part of the data
(not shown) exhibits qualitatively similar features. \label{fig:raw_data}}
\end{figure*}

\par\end{center}

\begin{center}
\begin{figure}
\begin{centering}
\includegraphics[scale=0.43]{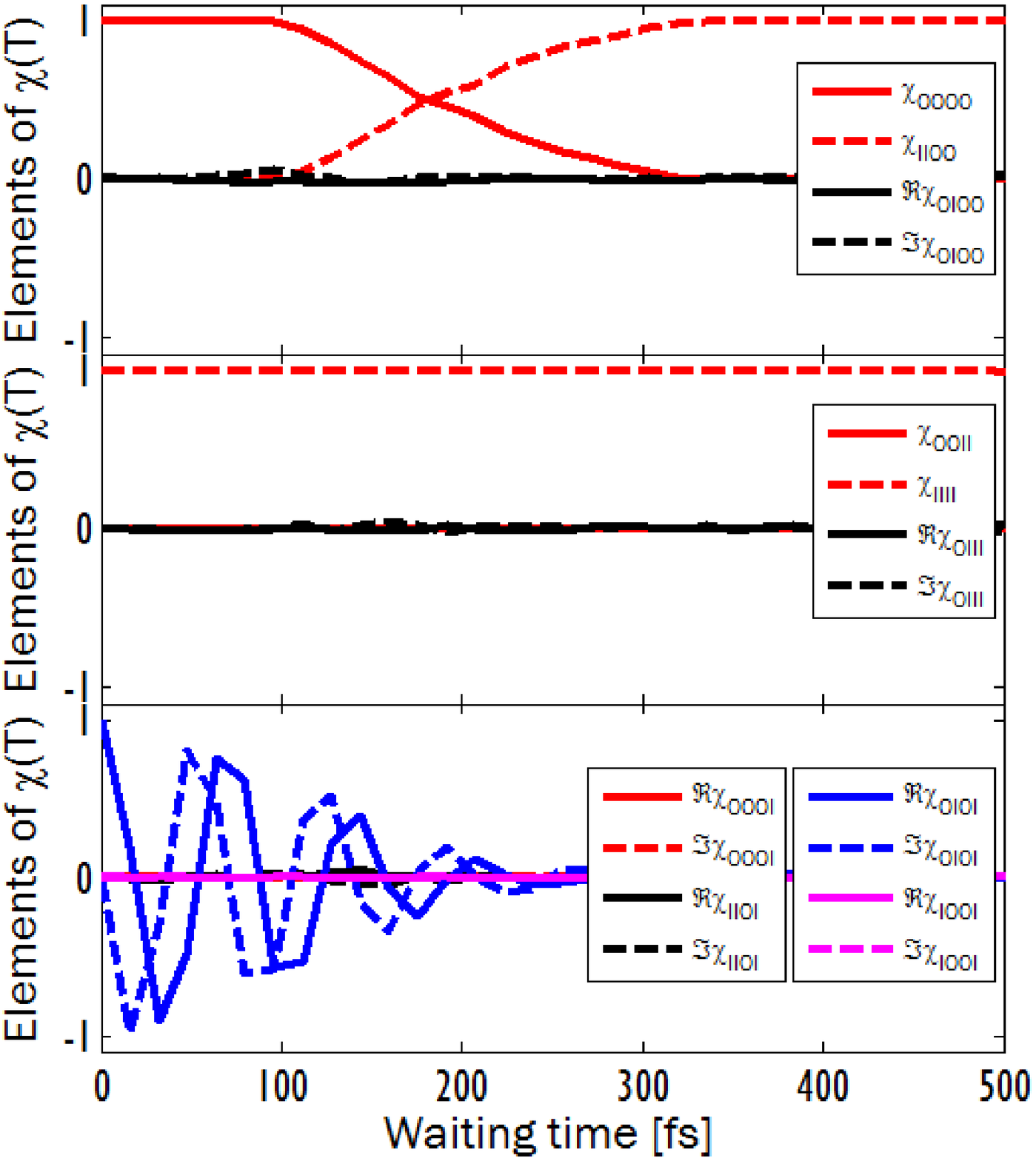}
\par\end{centering}

\caption{Nonzero elements of $\chi(T)$ extracted from the data of Fig. \ref{fig:raw_data}.
In the top panel, population transfer from the higher energy $|O\rangle$
to the lower energy $|I\rangle$ is monitored in the decay of $\chi_{OOOO}(T)$
and the rise of $\chi_{IIOO}(T)$ within the first 300 fs. The fall
of $\chi_{IIOO}(T)$ is presumably due to subsequent population decay
from $|I\rangle$ to $|g\rangle$. In the center panel, the fall from
$|I\rangle$ to $|g\rangle$ is observed via $\chi_{IIII}(T)$, although
uphill transfer to $|O\rangle$ is not observed from $\chi_{OOII}(T)$.
Finally, the right panel shows secular coherence dynamics that last
for about 150 fs, which indicates unexpected weak coupling between
the I and O states.\label{fig:tomo}}
\end{figure}

\par\end{center}

The result of this numerical procedure is in Fig. \ref{fig:tomo}.
Table 1 summarizes the values of the elements of $\chi(T)$ together
with their timescales given by fits with 95\% confidence intervals.
The full QPT analysis allows us to conclude that, in this system,
as anticipated, the non-secular terms $\chi_{IOOO}(T)$, $\chi_{IOII}(T)$,
$\chi_{IOOI}(T)$, $\chi_{IIOI}(T)$, and $\chi_{OOOI}(T)$ are negligible
throughout the first 500 fs, indicating weak coupling between populations
and coherences, as opposed to the situation of the Fenna-Matthews-Olson
complex \cite{Panitchayangkoon27122011}. On the other hand, $\chi_{OOOO}(T)$
and $\chi_{IIOO}(T)$ indicate that population from the higher $|O\rangle$
state transfers into $|I\rangle$ within 300 fs. The analogous situation
with $|I\rangle$ is different. Uphill transfer $|I\rangle\to|O\rangle$
is not observed, $\chi_{OOII}(T)\sim0$ throughout the experiment,
whereas population term $\chi_{IIII}(T)\sim1$ remains for all the
times of interest. Similarly, the explicitly monitored decay terms
$\chi_{ggqp}(T)$ are also negligible within that timescale, in consistency
with the reported timescales of radiative decay for supramolecular
aggregates (on the other of hundreds of picoseconds \cite{moll:6362}).
Similar conclusions were observed in pump-probe \cite{augulis} and
two-dimensional spectra on the system \cite{sperling}, although maybe
using a sample with a different morphology. Finally, we detect electronic
coherence between $|O\rangle$ and $|I\rangle$ which lasts for about
150-200 fs, allowing for a few quantum beats to occur before decoherence
sets in, indicating that the electronic coupling between the corresponding
localized exciton states is weak compared to the coupling of the electronic
states to the localized vibrational modes. This coupling was suggested
in \cite{sperling} in the form of weak cross-peaks of the two-dimensional
spectra, although quantum beats were not reported there, probably
due to a coarser sampling of the waiting time or to peak overlaps.
The weak coupling is also consistent with previous redox experiments
\cite{eisele}, and its decoherence timescale is similar to reported
values on a similar nanotube system with different chemical composition
\cite{Milota200945,moran-polarization-ii,moran-polarization,hauer}.
As shown in Table 1, the kinetics of the different processes in this
system are characterized by stretched exponentials with indices $\beta$
ranging between 1.6 and 2. We speculate that this is due to actual
exponential kinetics embedded in Gaussian disorder, but more studies
are needed to confirm this idea.

\begin{center}
\begin{figure*}
\centering{}%
\begin{tabular}{|c|c|c|}
\hline
\multicolumn{3}{|c}{Table 1. Summary of timescales of $\chi(T)$ }\tabularnewline
\hline
Process & Fit & Description\tabularnewline
\hline
\hline
$\chi_{OOOO}(T)\sim e^{-(T/\tau_{OO})^{\beta_{OO}}}$ & $\begin{array}{c}
\tau_{OO}=212\pm3\,\mbox{fs},\\
\beta_{OO}=3.3\pm0.2.
\end{array}$  & population decay\tabularnewline
\hline
$\chi_{IIOO}(T)\sim1-e^{-(T/\tau_{OO})^{\beta_{OO}}}$ & --- & population transfer\tabularnewline
\hline
$\chi_{IIII}(T)\sim1\,(>0.99)$ & --- & population decay\tabularnewline
\hline
$\chi_{OOII}(T)\sim0\,(<0.01)$ & --- & population transfer\tabularnewline
\hline
$\chi_{OIOI}(T)=\chi_{IOIO}^{*}(T)\sim e^{-i\bar{\omega}_{OI}T}e^{-(T/\tau_{OI})^{\beta_{OI}}}$ & $\begin{array}{c}
\frac{2\pi}{\bar{\omega}_{OI}}=70\pm4\,\mbox{fs},\\
\tau_{OI}=200\pm120\,\mbox{fs},\\
\beta_{OI}=2\pm1.
\end{array}$ & decoherence\tabularnewline
\hline
$\begin{array}{cc}
\chi_{IOOO}(T)=\chi_{OIOO}^{*}(T), & \chi_{IOII}(T)=\chi_{OIII}^{*}(T),\\
\mbox{\ensuremath{\chi}}_{IOOI}(T)=\chi_{OIIO}^{*}(T), & \mbox{\ensuremath{\chi}}_{IIOI}=\chi_{IIIO}^{*}(T),\\
 & \chi_{OOOI}(T)=\chi_{OOIO}^{*}(T)<0.08
\end{array}$ & --- & nonsecular terms\tabularnewline
\hline
\end{tabular}
\end{figure*}

\par\end{center}

\section*{Discussion}

We have demonstrated for the first time the realization of QPT on
a molecular system in condensed phase, namely, the inner and outer
wall excitons of a supramolecular light-harvesting nanotube. QPT has
been obtained through the collection of a series of frequency-resolved
TG spectra by systematically switching the frequency components of
the pulses at fixed polarization. Via numerical inversion of these
signals, we have reconstructed the full process matrix $\chi(T)$
for the dynamics of the excitons. We summarize the main qualitative
findings derived from the analysis of $\chi(T)$. First, an electronic
coherence between the inner and outer wall excitons persists for more
than a hundred femtoseconds, indicating a weak electronic coupling
between the excitons originating from different walls. Second, population
transfers quickly from the outer to the inner wall exciton within
the first hundreds of femtoseconds, but not the other way around.
These transfers deviate from simple exponential kinetics, although
this may be an effect of the ensemble measurements. Third, nonsecular
relaxation dynamics are measured to be negligible, suggesting that
the vibrational bath is dense and Markovian. These conclusions are
difficult to assess using a standard broadband approach, where these
processes are nontrivially convolved in a few peaks \cite{yuen-aspuru}.
Instead, our QPT protocol directly isolates each of these contributions
in a systematic way.

As we have shown, QPT can be easily carried out by a simple adaptation
of the traditional spectroscopic experiment to ensure that the maximum
amount of extractable information, at the quantum mechanical level,
is obtained. QPT can be interpreted as a procedure that reconstructs
the time-dependent quantum state of a system, and therefore, offers
a systematic and transparent way to design ultrafast spectroscopy
experiments. It complements the traditional approach where only specific
projections of the response of the material are collected. Therefore,
we envision many opportunities where the QPT paradigm will be powerful.
Specific examples include experiments on excitonic networks embedded
in complex environments in biological \cite{Panitchayangkoon27122011}
and solid state systems \cite{Stone05292009}, or reactive molecular
systems with strong vibronic features \cite{brixner_molecule} where
one expects an interesting interplay between electronic coherences
and populations beyond secular dynamics, and where the detailed imaging
of the quantum dynamics is required in order to construct theoretical
models. On the technical side, important directions will be the development
of compressed sensing approaches to ameliorate the scaling of QPT
as a function of system size \cite{shabaniprl,sanders}, or alternatively,
partial QPT protocols to pinpoint specific mechanisms that do not
require the knowledge of an entire process matrix $\chi(T)$. We foresee
exciting opportunities in which the QPT approach to ultrafast spectroscopy
will provide new insights into the excited state dynamics of chemical
systems.

\section*{Acknowledgements}

We are grateful for Prof. Marc Baldo's critical reading of the manuscript.
All the authors in this work were supported by the Center of Excitonics,
an Energy Frontier Research Center funded by the US Department of
Energy, Office of Science, Office of Basic Energy Sciences under Award
Number DESC0001088. In addition, D.M.E. was partially supported by
the Feodor Lynen Research Fellowship from the Alexander von Humboldt-Foundation,
J.J.K. was also supported by the Harvard University Center for the
Environment and NSERC, and C.P.S. was also supported by an NSF Graduate
Research Fellowship.

\bibliographystyle{unsrt}

\end{document}


\begin{center}
\setcounter{page}{1}
\par\end{center}

\begin{center}
{\huge{Supplementary information}}
\par\end{center}{\huge \par}

\global\long\def\theequation{S\arabic{equation}}
 \setcounter{equation}{0}

\global\long\def\thefigure{S\arabic{figure}}
 \setcounter{figure}{0}

\section{Synthesis of individual supramolecular light-harvesting nanotubes
in solution}

The amphiphilic cyanine dye derivative 3,3\textasciiacute{}-bis(2-sulfopropyl)-5,5\textasciiacute{},6,6\textasciiacute{}-tetrachloro-1,1\textasciiacute{}-dioctylbenzimidacarbocyanine
(C8S3, MW=902.8 g mol$^{-1}$, Fig. 1 in main text) was obtained as
a sodium salt (FEW Chemicals) and used as received. The individual
supramolecular light-harvesting nanotubes, consisting of concentric
walls of excitons, were prepared in water/methanol as described in
\cite{eiselenanotech}. Solutions of nanotubes were stored in the
dark and used for experiments within four hours. Absorption spectroscopy
was used as a tool to monitor the aggregation process before and during
the non-linear spectroscopy experiments. We limited our investigation
to samples that contained the expected spectral contributions from
individual supramolecular light-harvesting nanotubes and had no significant
spectral contributions from bundled single-walled light-harvesting
nanotubes \cite{eisele}.

\section{Description of optical setup}

A non-collinear parametric amplifier \cite{cerullo} (NOPA) is pumped
by a regeneratively amplified Ti:sapphire laser at 800 nm with a pulse
energy of 350 \textgreek{m}J at a repetition rate of 10 kHz. The NOPA
produces pulses with a central frequency of 505 THz, full-width at
half-maximum of 21 THz, and approximately equal intensities at 501
and 510 THz, i.e., the energies of the $|I\rangle$ and the $|O\rangle$
states. The pulses are compressed with a prism pair to approximately
20-25 fs.

After the NOPA, the beam passes through a 2D phase mask optimized
for first-order diffraction to produce four beams in the BOXCARS geometry.
The beams then enter a diffraction-based pulse shaper using a Hamamatsu
X7550 2D spatial light modulator (SLM) for phase and amplitude shaping
of the frequency components of each beam \cite{colbert}.\textsuperscript{}The
beams are spectrally dispersed by a grating and imaged at different
vertical positions by a cylindrical lens onto the SLM for independent
temporal shaping. We apply a sawtooth grating pattern in the vertical
dimension of the SLM device, enabling the amplitude of the frequency
components of each beam to be controlled by the amplitude of the grating.
In the experiments, a Gaussian amplitude filter is applied via SLM
to each beam in order to diffract only the frequencies covering a
single transition. Each beam has approximately 3.5 nJ/pulse for the
broadband spectrum and 450-500 pJ/pulse for the narrowband spectra.

After pulse shaping, the beams are imaged onto the sample to perform
a transient grating experiment. The first two narrowband pulses, with
wavevectors $\boldsymbol{k}_{1}$ and $\boldsymbol{k}_{2}$, generate
a spatially periodic excitation grating in the material due to the
change in the refractive index upon excitation. The system is probed
by the third narrowband pulse after a time delay. The third pulse,
with wavevector $\boldsymbol{k}_{3}$, diffracts off the grating into
the TG direction, $\boldsymbol{k}_{s}=-\boldsymbol{k}_{1}+\boldsymbol{k}_{2}+\boldsymbol{k}_{3}$.
The signal co-propagates with the fourth beam, which acts as a (broadband)
local oscillator for heterodyne-detection. Spectral interferometry
is used to retrieve both the real and imaginary parts of the signal.

This setup is used to obtain 8 different frequency-resolved TG spectra
where the first three pulses are narrowband and selective to specific
transitions. Between the collection of each TG spectrum, a linear
absorption measurement is obtained to ensure that the sample does
not degrade.

\section{TG experiment as a QPT\label{sub:TG-experiment-as}}

The basic idea of carrying out QPT using information from eight different
TG spectra collected in the experiment has been intuitively explained
in the main text. Here we elaborate on the formal details.

The three pulses interacting with the sample have carrier (center)
frequencies $\omega_{1},\omega_{2},\omega_{3}$ which are close to
the transition energies $\omega_{Ig}=\omega_{II,I}=\omega_{IO,O}$
or $\omega_{Og}=\omega_{OO,O}=\omega_{IO,I}$. We label the first,
second, and third pulses as $p,q,r=\mbox{I},\mbox{O}$, respectively,
depending on whether they are centered close to $\omega_{Ig}$ or
$\omega_{Og}$. The pulses generate a third order time-dependent polarization
which is detected by the LO pulse (fourth pulse) which, for our purposes,
is ideally broadband, covering all the transitions of interest. Under
this condition, the complex-valued \emph{frequency-resolved} TG spectrum
as a function of waiting time $T$ and frequency $\omega$ can be
immediately related to the half-sided Fourier transform of the complex-valued
TG polarization $P_{\boldsymbol{k}_{s}}^{pqr}(\tau=0,T,t)$%
\footnote{More precisely, the scalar $P_{\boldsymbol{k}_{s}}^{pqr}$ is the
projection of the TG polarization on the LO field, which is parallel
to the axes of the nanotubes (as are the other pulses). Note the dual
use of the word \emph{polarization. }Here, it refers to the electric
dipole density in the material, but we also use it to denote the direction
of the electric field of the pulses.%
} via,

\begin{equation}
[S_{TG}]^{pqr}(\omega,T)=\int_{0}^{\infty}dte^{i\omega t}P_{\boldsymbol{k}_{s}}^{pqr}(\tau=0,T,t).\label{eq:S_TG}
\end{equation}
Here, $\tau$ (coherence time) and $T$ (waiting time) correspond
to the time intervals between the first and the second, and the second
and the third pulses, respectively%
\footnote{When $\tau\neq0$, the TG experiment generalizes to the photon-echo
configuration, which for our QPT purposes is not necessary.%
}. The free-induction decay time of the TG polarization is associated
with $t$ (sometimes known as echo-time). Since the half-sided Fourier
transform in Eq. (\ref{eq:S_TG}) is associated with this time interval,
the set of emission frequencies in the TG spectrum $[S_{TG}]^{pqr}(T,\omega)$
corresponds to this free-induction decay. These frequencies are associated
with the optical coherences between $|g\rangle$ and the SEM, or between
the SEM and the DEM, and they correlate with the detection of different
populations and coherences by the end of the waiting time. Consider
the scenario where dissipative processes of these optical coherences
are not spectrally broader than the separation between the different
peaks in the TG spectra, which is what happens in our case. Then,
for purposes of QPT, one can properly define the integrated amplitude
of the spectra across a specific spectral window of width $2\sigma_{4}\equiv330\,\mbox{cm}^{-1}$
about the peak centered at a particular frequency $\omega_{4}$%
\footnote{The peaks are separated from one another by $\sim$330 $\mbox{cm}$$^{-1}$
so windows of width 330 $\mbox{cm}^{-1}$ are sufficiently narrow
to detect single peak contributions, and broad enough to contain all
the relevant spectral diffusion. %
},

\begin{eqnarray}
[\bar{S}_{TG}]^{pqr}(\omega_{4},T) & \equiv & \int_{\omega_{4}-\sigma_{4}}^{\omega_{4}+\sigma_{4}}d\omega[S_{TG}]^{pqr}(\omega,T)\nonumber \\
 & = & \int_{-\infty}^{\infty}dt\underbrace{\Theta(t)\sigma_{4}\mbox{sinc}\sigma_{4}te^{i\omega_{4}t}}_{\equiv E_{4}^{*}(t)}P_{\boldsymbol{k}_{s}}^{pqr}(\tau=0,T,t),\label{eq:frequency_integrate}
\end{eqnarray}
where we have used the step function $\Theta(t)$. The interpretation
of Eq. (\ref{eq:frequency_integrate}) is quite intuitive and reads
as follows: Integrating the (broadband LO) \emph{frequency-resolved
}complex amplitude $[S_{TG}]^{pqr}(T,\omega)$ across a spectral window
$\omega\in[\omega_{4}-\sigma_{4},\omega_{4}+\sigma_{4}]$ is equivalent
to collecting the total TG photon-count signal arising from the overlap
between an \emph{effectively narrowband} LO pulse $E_{4}(t)$ (with
carrier frequency $\omega_{4}$ and time-width $\sim\sigma_{4}^{-1}$)
centered at the end of the waiting time (at the same time as the third
pulse $r$, at $t=0$) and the $t$ dependent TG polarization $P_{\boldsymbol{k}_{s}}^{pqr}$
undergoing free-induction decay. $\omega_{4}$ is chosen to be resonant
with one of the emission frequencies. $E_{4}$ is short in time (impulsive,
broadband), meaning that $\sigma_{4}$ is wide enough to cover the
dynamic broadening of a given optical transition. Yet, it is long
in time (narrowband) enough to only be selective with respect to the
different transitions. In previous articles, we have shown that a
TG signal with four ``impulsive-yet-selective'' pulses prepares
and detects populations and coherences in the SEM via the first two
and the last two pulses in such a way that the TG experiment may be
regarded as a QPT experiment. Hence, from Eq. (\ref{eq:frequency_integrate}),
we conclude that QPT can also be achieved via the frequency-resolved
TG spectra in this article \cite{yuen-aspuru,yuenzhou}.

Fig. 3 in main text shows that the possible emission frequencies,
and hence values of $\omega_{4}$, in the different TG spectra are
dictated by the third pulse $r$. If $r=O$, the induced TG optical
coherence and therefore $\omega_{4}$ take values close to $\omega_{Ig}=\omega_{II,I}=\omega_{IO,O}$
via SE and ESA, at $\omega_{Og}=\omega_{OO,O}=\omega_{IO,I}$ via
GSB, SE, ESA, or GSR, or at $\omega_{OO,I}$ via ESA. Similarly, if
$r=I$, $\omega_{4}$ can take values close to $\omega_{II,O}$ via
ESA, to $\omega_{Ig}=\omega_{II,I}=\omega_{IO,O}$ via GSB, SE ESA,
or GSR, and $\omega_{Og}=\omega_{IO,I}$ via SE or ESA. Hence, for
each of the eight frequency-resolved TG spectra $[S_{TG}]^{pqr}(\omega,T)$,
there are three possible ``carrier frequencies'' $\omega_{4}$ from
which we can extract TG signals $[\bar{S}_{TG}]^{pqr}(\omega_{4},T)$,
yielding a total of 24 complex numbers as a function of $T$.

One can readily obtain explicit expressions for $[\bar{S}_{TG}]^{pqr}(\omega_{4},T)$
by translating the double-sided Feynman diagrams in Fig. 3 in main
text in terms of the initial states prepared by the first two pulses,
and the final states detected by the last two pulses \cite{yuenzhou,witness,joel_chapter}.
If $r=O$, these are,

\begin{eqnarray}
[\bar{S}_{TG}]^{pqO}(\omega_{4},T) & = & C_{pqO}\overbrace{\mu_{pq}\mu_{qg}}^{\mbox{initial state preparation}}\nonumber \\
 &  & \times\begin{cases}
\underbrace{\overbrace{\mu_{Og}\mu_{Ig}}^{\mbox{final state detection}}\chi_{IOqp}(T)}_{SE}\underbrace{-\mu_{IO,I}\mu_{IO,O}\chi_{IOqp}(T)}_{ESA} & \mbox{for \ensuremath{\omega_{4}=\omega_{Ig}}},\\
\underbrace{\mu_{Og}^{2}\delta_{qp}}_{GSB}\underbrace{-\mu_{Og}^{2}\chi_{ggqp}(T)}_{GSR}\underbrace{+\mu_{Og}^{2}\chi_{OOqp}(T)}_{SE}\underbrace{-\mu_{OO,O}^{2}\chi_{OOqp}(T)}_{ESA}\underbrace{-\mu_{IO,I}^{2}\chi_{IIqp}(T)}_{ESA} & \mbox{for \ensuremath{\omega_{4}=\omega_{Og}}},\\
\underbrace{-\mu_{OO,O}\mu_{OO,I}\chi_{OIqp}(T)}_{ESA} & \mbox{for \ensuremath{\omega_{4}=\omega_{OO,I}}},
\end{cases}\label{eq:S_TG_pqO}
\end{eqnarray}
and the analogous expressions hold for $[\bar{S}_{TG}]^{pqI}(\omega_{4},T)$
upon the substitutions$O\to I$ and $OO\to II$. Here, we have highlighted
the dipole transitions $\mu_{ij}$ associated with the initial state
preparation and the final state detection in each case. We have also
assumed that $\mu_{ij}=\mu_{ji}$ since the excitonic states can be
taken to be real due to time-reversal symmetry. For the $\omega_{4}=\omega_{Og}$
case, it is possible to simplify the expression by assuming that the
total exciton population during the waiting time is distributed exclusively
among $|O\rangle$, $|I\rangle$, and $|g\rangle$,

\begin{equation}
\chi_{OOqp}(T)+\chi_{IIqp}(T)+\chi_{ggqp}(T)=\delta_{qp},\label{eq:norm_preservation}
\end{equation}
so that it reads,

\begin{equation}
[\bar{S}_{TG}]^{pqO}(\omega_{4},T)=C_{pqO}\mu_{pg}\mu_{qg}(2\mu_{Og}^{2}-\mu_{OO,O}^{2})\chi_{OOqp}(T)+(\mu_{Og}^{2}-\mu_{IO,I}^{2})\chi_{IIqp}(T)\,\,\,\mbox{for \ensuremath{\omega_{4}=\omega_{Og}}}.\label{eq:S_TG_pqO_trace}
\end{equation}
This approximation relies on two assumptions: (a) That there are no
uphill transfers of population to the DEM during the waiting time,
which is very reasonable considering the large energy gap between
the SEM and the DEM, and (b) that the transfer to the dark states
is also negligible.

$C_{pqO}$ indicates the joint transition probability amplitude to
carry out the three different dipole transitions via the three different
pulses. Whereas in principle one can obtain explicit expressions for
this amplitude, in the present case, the narrowband pulses with imperfect
Gaussian forms, the pulse overlaps, as well as the broadening of the
TG transitions due to dynamic disorder altogether impede its precise
determination. We shall write it as,

\begin{eqnarray}
C_{pqO} & = & f_{pq}E_{p}(\omega_{pg})E_{q}(\omega_{qg})E_{O}(\omega_{Og})\nonumber \\
 & \approx & f_{pq}\mbox{max}(E_{p}(\omega))\mbox{max}(E_{q}(\omega))\mbox{max}(E_{O}(\omega)).\label{eq:max}
\end{eqnarray}
Here, we have used the fact that the pulses are narrowband and centered
about the relevant transitions ($E_{p}(\omega_{pq})\approx\mbox{max}(E_{p}(\omega))$
and so on), and we extract the respective amplitudes from the power
spectra of the pulses, $E_{i}(\omega)=\sqrt{|E_{i}(\omega)|^{2}}$
(assuming $E_{i}(\omega)$ has no chirp and its global phase is already
considered in the phasing procedure with respect to the other pulses).
We hide all the complexity of $C_{pqO}$ in the complex-valued factor
$f_{pq}$ which takes into account the overlap between pulses $p$
and $q$. Finally, from the absorption spectrum, we can get a good
estimate of
\begin{equation}
\frac{\mu_{Og}}{\mu_{Ig}}\approx\sqrt{\frac{A(\omega_{Og})}{A(\omega_{Ig})}},\label{eq:estimate_mu}
\end{equation}
where $A(\omega)$ is the absorption spectrum of the material. Note
that the contributions corresponding to SE/GSB and ESA/GSR involve
a net gain and loss of photons to the electric field in the $\boldsymbol{k}_{s}$
direction, respectively, and hence come with opposite signs. Also,
GSB appears only if the first two pulses are resonant with the same
transitions and therefore create a population (rather than a coherence)
in the excited state, and hence, it is proportional to $\delta_{pq}$.
Since the GSB term monitors (stationary) ground state population during
the waiting time $T$, it is proportional to $\chi_{gggg}(T)=1$ and
shows up as a $T$-independent background%
\footnote{This is not true if the pump pulses 1 and 2 prepare a nuclear wavepacket
in $|g\rangle$ which is different from its initial equilibrium configuration
\cite{banin}. This will not happen as long as the impulsive limit
is satisfied and the Condon approximation is valid. %
}.

So far, we have 24 effective narrowband time (or frequency) integrated
complex-valued TG signals $[\bar{S}_{TG}]^{pqr}(T,\omega)$ which
amount to 48 real-valued data points as a function of $T$. Note that
in general, these signals are linear combinations of different elements
of $\chi(T)$ and, in fact, according to Eq. (\ref{eq:S_TG_pqO}),
several signals report on a single element of $\chi(T)$ at a time.
Let us now count the number of elements of $\chi(T)$ to invert for
our two-level system composed of $|I\rangle$ and $|O\rangle$. Hermicity
of $\chi(T)$ requires that $\chi_{ijqp}(T)=\chi_{jipq}^{*}(T)$.
This amounts to the real-valued population terms $\chi_{OOOO}(T)$,
$\chi_{IIOO}(T)$, $\chi_{IIII}(T)$, and $\chi_{OOII}(T)$, and the
complex-valued $\chi_{IOIO}(T)=\chi_{OIOI}^{*}(T)$, together with
the non-secular (not energy conserving, also complex-valued) terms
$\chi_{IOOO}(T)=\chi_{OIOO}^{*}(T)$, $\chi_{IOII}(T)=\chi_{IOII}(T)$,
$\chi_{IOOI}(T)=\chi_{OIIO}^{*}(T)$, $\chi_{OOIO}(T)=\chi_{OOOI}^{*}(T)$,
and $\chi_{IIIO}(T)=\chi_{IIOI}^{*}(T)$. Based on this symmetry,
there are 16 real parameters of $\chi(T)$ to extract %
\footnote{Had the population stayed only in $|I\rangle$ and $|O\rangle$ (and
not in $|g\rangle$ via GSR), the number of unknowns would have been
reduced to 12 real parameters \cite{yuenzhou}.%
} out of a redundant set of 48 real-valued data points.

\section{Energy level assignments}

Energies of the SEM and DEM states addressed in our experiment have
been self-consistently assigned from the frequency-resolved TG spectra.
As a first examination, from the linear absorption, peak maxima corresponding
to $|I\rangle$ and $|O\rangle$ are located at $\omega_{Ig}=16695\,\mbox{cm}^{-1}$
and $\omega_{Og}=16970\,\mbox{cm}^{-1}$, respectively. These peaks
are broadened both by static and dynamic disorder of the ensemble.
As shown in Fig. 2 in main text, narrowband excitation in the experiment
is effected in such a way that the pulses are centered at the edge
of each band, therefore selecting only a subset of realizations of
static disorder. Therefore, the average energies in the linear absorption
do not coincide with those probed in the TG experiment. Hence, it
is more accurate to extract the energy levels of interest from the
TG spectra themselves using the initial condition
\begin{equation}
\chi_{ijqp}(0)=\delta_{iq}\delta_{jp},\label{eq:initial_condition}
\end{equation}
For instance, whereas the OOO spectrum can potentially contain three
different emission frequencies, at $T=0$ it consists of a single
peak%
\footnote{In fact, as we shall explain later, the two other potential peaks
have negligible amplitudes even at $T>0$.%
} with maximum amplitude at $\omega\sim17068\,\mbox{cm}^{-1}$. This
peak must correspond to (see Eq. (\ref{eq:S_TG_pqO}); also Fig. 3
in main text, left top panel) $\chi_{OOOO}(0)=1$, in a combination
of SE, ESA, and GSB processes. Whereas SE/GSB is expected to show
up at $\sim$3.5 $\mbox{cm}^{-1}$ red-shifted from ESA at cryogenic
temperatures \cite{augulis}, dynamic and some static disorder at
room temperature forbids an unambiguous discrimination as it broadens
peaks up to a total width of about 330 $\mbox{cm}^{-1}$, as mentioned
at the beginning of SI, Sec. \ref{sub:TG-experiment-as}. From here,
we infer that $\omega_{Og},\omega_{OO,O}\sim17068$ $\mbox{cm}^{-1}$.
Analogously, from the III spectrum at $T=0$ and $\chi_{IIII}(0)=1$,
we obtain $\omega_{Ig},\omega_{II,I}\sim16635\,\mbox{cm}^{-1}$. Based
on these observations, we use $\omega_{Og}=\omega_{OO,O}=17068\,\mbox{cm}^{-1}$
and $\omega_{Ig}=\omega_{II,I}=16635\,\mbox{cm}^{-1}$.

The presence of the SEM states $|I\rangle$ and $|O\rangle$ demand
the consideration of an additional combination exciton $|IO\rangle$,
which we treat as a doubly-excited state where the two excitons are
present, one in $|I\rangle$ and the other in $|O\rangle$, and its
energy is the sum of the two SEM exciton energies, $\omega_{IO,O}=\omega_{Ig}$
and $\omega_{IO,I}=\omega_{Og}$. This is a reasonable assumption
considering that the interactions between the $|I\rangle$ and the
$|O\rangle$ excitons will be weak across the 4 nm hydrophobic core
separating them.

We confirm the extracted energies by analyzing the rest of the TG
spectra at $T=0$. First, OOI and IIO spectra each contain a single
peak at 16572 and 17025 $\mbox{cm}^{-1}$, respectively. Due to the
frequencies of the pulses involved in these two experiments, only
GSB and ESA processes contribute at $\omega=\omega_{Ig}=\omega_{IO,O}$
and $\omega=\omega_{Og}=\omega_{IO,I}$, which is to a good approximation
what we see. Second, spectra IOI and OIO show peaks at 17012 and 16635
$\mbox{cm}^{-1}$, associated with SE and ESA at $\omega=\omega_{Og}=\omega_{IO,I}$
and $\omega=\omega_{Ig}=\omega_{IO,O}$. Finally, IOO and OII spectra
show peaks at 17452 and 16118 $\mbox{cm}^{-1}$ corresponding to ESA
at $\omega=\omega_{OO,I}$ and $\omega=\omega_{II,O}$. These observations
validate the energy assignments in Fig. 2b in main text.

\section{Data processing}

As explained in SI Sec. \ref{sub:TG-experiment-as}, Fig. 3 in main
text and Eq. (\ref{eq:S_TG_pqO}) comprehensively enumerate the possible
processes within the SEM that can be detected from the eight different
TG spectra beyond $T=0$. In principle, they manifest as three spectrally
well-separated peaks in each TG spectrum, indicating general transfers
amongst populations and coherences.

For each of the possible TG emission frequencies $\omega_{4}$, we
have computed the integral of the raw complex spectra given by Eq.
(\ref{eq:frequency_integrate}) using a half-width of $\sigma_{4}=165\,\mbox{cm}^{-1}$.
Since the centers of the bands are separated farther than 330 $\mbox{cm}^{-1}$
from one another, the TG emission bands are very well-separated. The
obtained set of signals is quite sparse. Table S1 shows the normalized
contribution of $\sum_{T}|[\bar{S}_{TG}]^{pqr}(\omega_{4},T)|^{2}$
for each frequency-resolved TG spectrum. Together with each entry,
we have also indicated the element of $\chi(T)$ associated with each
signal. For instance, the peak centered at $\omega_{Og}=\omega_{OO,O}$
in the IIO spectra reports on both $\chi_{OOII}(T)$ and $\chi_{IIII}(T)$,
whereas the peak at $\omega_{II,O}$ in OII is directly proportional
to $\chi_{IOIO}$. To obtain a rough idea of the experimental data,
we have highlighted the entries that contribute the most per TG spectrum,
and most of them account for over 97\% of the total norm of the respective
experiment, yielding what looks like a sparse data set.

\begin{center}
\begin{tabular}{|c|c|c|c|}
\hline
\multicolumn{4}{|c}{TABLE S1. Normalized contribution of $\sum_{T}|[\bar{S}_{TG}]^{pqr}(\omega_{4},T)|^{2}$}\tabularnewline
\hline
TG spectrum\textbackslash{}$\omega_{4}$ $[\mbox{cm}^{-1}]$ & $\omega_{Ig}=16635\,\mbox{cm}^{-1}$ & $\omega_{Og}=\omega_{OO,O}=\omega_{IO,I}=17068\,\mbox{cm}^{-1}$ & $\omega_{OO,I}=17501\,\mbox{cm}^{-1}$\tabularnewline
\hline
\hline
OOO & 0.0001 ($\chi_{IOOO}$)  & \textbf{0.9999} (\textbf{$\chi_{OOOO}$}, $\chi_{IIOO}$)  & 0.0000 $(\chi_{OIOO})$\tabularnewline
\hline
IIO & 0.0622 ($\chi_{IOII}$) & \textbf{0.9007} (\textbf{$\chi_{OOII}$}, $\chi_{IIII}$)  & 0.0371 $(\chi_{OIII})$\tabularnewline
\hline
IOO & 0.0320 ($\chi_{IOOI}$) & 0.1231 ($\chi_{OOOI}$, $\chi_{IIOI}$)  & \textbf{0.8449} $(\chi_{OIOI})$\tabularnewline
\hline
OIO & \textbf{0.9973 }($\chi_{IOIO}$) & 0.0018 ($\chi_{OOIO}$, $\chi_{IIIO}$)  & 0.0009 $(\chi_{OIIO})$\tabularnewline
\hline
\hline
TG spectrum\textbackslash{}$\omega_{4}$ $[\mbox{cm}^{-1}]$ & $\omega_{II,O}=16202\,\mbox{cm}^{-1}$ & $\omega_{Ig}=\omega_{II,I}=\omega_{IO,O}=16635\,\mbox{cm}^{-1}$ & $\omega_{Og}=17068\,\mbox{cm}^{-1}$\tabularnewline
\hline
OOI & 0.0221 ($\chi_{IOOO}$) & \textbf{0.97}7\textbf{9} (\textbf{$\chi_{OOOO}$}, $\chi_{IIOO}$) & 0.0000 ($\chi_{OIOO}$)\tabularnewline
\hline
III & 0.0050 ($\chi_{IOII}$)  & \textbf{0.9947 }(\textbf{$\chi_{OOII}$}, $\chi_{IIII}$) & 0.0003 ($\chi_{OIII}$)\tabularnewline
\hline
IOI & 0.0018 ($\chi_{IOOI}$)  & 0.0689 ($\chi_{IIOI}$, $\chi_{OOOI}$) & \textbf{0.9294 }($\chi_{OIOI}$)\tabularnewline
\hline
OII & \textbf{0.9886} ($\chi_{IOIO}$)  & 0.0061 ($\chi_{IIIO}$, $\chi_{OOIO}$) & 0.0052 ($\chi_{OIIO}$)\tabularnewline
\hline
\end{tabular}
\par\end{center}

Note that the entries with small contributions correspond to nonsecular
terms. Whereas this table serves as an illustration to the rationale
behind our procedure, we do not use it for the numerics \textit{per
se}, as the signals do not correspond to the elements of $\chi(T)$
alone, but are weighed by dipole moment and electric field terms.
To proceed in a more systematic fashion, we follow the following procedure:
\begin{enumerate}
\item From each signal $[\bar{S}_{TG}]^{pqr}(\omega_{4},T)$ in Eq. (\ref{eq:S_TG_pqO}),
construct

\begin{equation}
[\bar{s}_{TG}]^{pqr}(\omega_{4},T)=\frac{[\bar{S}_{TG}]^{pqr}(\omega_{4},T)}{\mbox{max}(E_{p}(\omega))\mbox{max}(E_{q}(\omega))\mbox{max}(E_{r}(\omega))\mu_{pq}\mu_{gq}},\label{eq:little_S_TG}
\end{equation}

where the dipoles are given in units of $\mu_{Ig}$ (using Eq. (\ref{eq:estimate_mu})).

\item Taking into account the initial condition Eq. (\ref{eq:initial_condition})
in Eqs. (\ref{eq:little_S_TG}) and (\ref{eq:S_TG_pqO_trace}) as
well as their analogues upon the $O\to I$ and $OO\to II$ substitutions,
yields the following coefficients,

\begin{eqnarray}
A & \equiv & [\bar{s}_{TG}]^{OOO}(\omega_{Og},0)=f_{OO}(2\mu_{Og}^{2}-\mu_{OO,O}^{2}),\nonumber \\
B & \equiv & [\bar{s}_{TG}]^{IIO}(\omega_{Og},0)=f_{II}(\mu_{Og}^{2}-\mu_{IO,I}^{2}),\nonumber \\
C & \equiv & [\bar{s}_{TG}]^{OOI}(\omega_{Ig},0)=f_{OO}(\mu_{Ig}^{2}-\mu_{IO,O}^{2}),\nonumber \\
D & \equiv & [\bar{s}_{TG}]^{III}(\omega_{Ig},0)=f_{II}(2\mu_{Ig}^{2}-\mu_{II,I}^{2}),\nonumber \\
E & \equiv & [\bar{s}_{TG}]^{IOI}(\omega_{Og},0)=f_{IO}(\mu_{Ig}\mu_{Og}-\mu_{IO,I}\mu_{IO,O}),\nonumber \\
F & \equiv & [\bar{s}_{TG}]^{IOO}(\omega_{OO,I},0)=f_{IO}(\mu_{OO,O}\mu_{OO,I}),\nonumber \\
G & \equiv & [\bar{s}_{TG}]^{OIO}(\omega_{Ig},0)=f_{OI}(\mu_{Og}\mu_{Ig}-\mu_{IO,O}\mu_{IO,I}).\nonumber \\
H & \equiv & [\bar{s}_{TG}]^{OII}(\omega_{II,O},0)=f_{OI}(\mu_{II,I}\mu_{II,O}).\label{eq:little_STG_initial}
\end{eqnarray}

These coefficients precisely constitute the set of dipole combinations
required for the inversion of $\chi(T)$ from Eq. (\ref{eq:S_TG_pqO})).

\item In order to make use of linear algebra, the coefficients from Eq.
(\ref{eq:little_STG_initial}) are arranged into matrices,

\begin{equation}
\underbar{\ensuremath{\mathbb{M}}}_{OO}=\underbar{\ensuremath{\mathbb{M}}}{}_{II}\equiv\left[\begin{array}{cccc}
0 & 0 & G & -iG\\
A & B & 0 & 0\\
0 & 0 & F & iF\\
0 & 0 & H & -iH\\
C & D & 0 & 0\\
0 & 0 & E & iE
\end{array}\right],\,\,\underbar{\ensuremath{\mathbb{M}}}_{OI}\equiv\left[\begin{array}{cccccccc}
0 & 0 & G & 0 & 0 & 0 & -iG & 0\\
A & B & 0 & 0 & -iA & -iB & 0 & 0\\
0 & 0 & 0 & F & 0 & 0 & 0 & -iF\\
0 & 0 & H & 0 & 0 & 0 & -iH & 0\\
C & D & 0 & 0 & -iC & -iD & 0 & 0\\
0 & 0 & 0 & E & 0 & 0 & 0 & -iE\\
0 & 0 & 0 & G & 0 & 0 & 0 & iG\\
A & B & 0 & 0 & iA & iB & 0 & 0\\
0 & 0 & F & 0 & 0 & 0 & iF & 0\\
0 & 0 & 0 & H & 0 & 0 & 0 & iH\\
C & D & 0 & 0 & iC & iD & 0 & 0\\
0 & 0 & E & 0 & 0 & 0 & iE & 0
\end{array}\right],\label{eq:matrix_M}
\end{equation}

whereas the signals are organized as vectors,

\begin{equation}
\mathbb{S}_{OO}(T)=\left[\begin{array}{c}
[\bar{s}_{TG}]^{OOO}(\omega_{Ig},T)\\
{}[\bar{s}_{TG}]^{OOO}(\omega_{Og},T)\\
{}[\bar{s}_{TG}]^{OOO}(\omega_{OO,I},T)\\
{}[\bar{s}_{TG}]^{OOI}(\omega_{II,O},T)\\
{}[\bar{s}_{TG}]^{OOI}(\omega_{Ig},T)\\
{}[\bar{s}_{TG}]^{OOI}(\omega_{Og},T)
\end{array}\right],\,\,\mathbb{S}_{II}(T)=\left[\begin{array}{c}
[\bar{s}_{TG}]^{IIO}(\omega_{Ig},T)\\
{}[\bar{s}_{TG}]^{IIO}(\omega_{Og},T)\\
{}[\bar{s}_{TG}]^{IIO}(\omega_{OO,I},T)\\
{}[\bar{s}_{TG}]^{III}(\omega_{II,O},T)\\
{}[\bar{s}_{TG}]^{III}(\omega_{Ig},T)\\
{}[\bar{s}_{TG}]^{III}(\omega_{Og},T)
\end{array}\right],\,\,\mathbb{S}_{OI}(T)=\left[\begin{array}{c}
[\bar{s}_{TG}]^{OIO}(\omega_{Ig},T)\\
{}[\bar{s}_{TG}]^{OIO}(\omega_{Og},T)\\
{}[\bar{s}_{TG}]^{OIO}(\omega_{OO,I},T)\\
{}[\bar{s}_{TG}]^{OII}(\omega_{II,O},T)\\
{}[\bar{s}_{TG}]^{OII}(\omega_{Ig},T)\\
{}[\bar{s}_{TG}]^{OII}(\omega_{Og},T)\\
{}[\bar{s}_{TG}]^{IOO}(\omega_{Ig},T)\\
{}[\bar{s}_{TG}]^{IOO}(\omega_{Og},T)\\
{}[\bar{s}_{TG}]^{IOO}(\omega_{OO,I},T)\\
{}[\bar{s}_{TG}]^{IOI}(\omega_{II,O},T)\\
{}[\bar{s}_{TG}]^{IOI}(\omega_{Ig},T)\\
{}[\bar{s}_{TG}]^{IOI}(\omega_{Og},T)
\end{array}\right].\label{eq:signals_S}
\end{equation}

The goal is to extract $\chi(T)$, which is also written as a series
of vectors,

\[
\mathbb{X}_{OO}(T)\equiv\left[\begin{array}{c}
\chi_{OOOO}(T)\\
\chi_{IIOO}(T)\\
\Re\{\chi_{OIOO}(T)\}\\
\Im\{\chi_{OIOO}(T)\}
\end{array}\right],\,\,\mathbb{X}_{II}(T)\equiv\left[\begin{array}{c}
\chi_{OOII}(T)\\
\chi_{IIII}(T)\\
\Re\{\chi_{OIII}(T)\}\\
\Im\{\chi_{OIII}(T)\}
\end{array}\right],\,\,\mathbb{X}_{OI}(T)=\left[\begin{array}{c}
\Re\{\chi_{OOOI}\}\\
\Re\{\chi_{IIOI}\}\\
\Re\{\chi_{OIOI}\}\\
\Re\{\chi_{IOOI}\}\\
\Im\{\chi_{OOOI}\}\\
\Im\{\chi_{IIOI}\}\\
\Im\{\chi_{OIOI}\}\\
\Im\{\chi_{IOOI}\}
\end{array}\right],
\]

which fulfill,

\begin{eqnarray}
\underbar{\ensuremath{\mathbb{M}}}_{OO}\mathbb{X}_{OO}(T) & = & \mathbb{S}_{OO}(T),\nonumber \\
\underbar{\ensuremath{\mathbb{M}}}_{II}\mathbb{X}_{II}(T) & = & \mathbb{S}_{II}(T),\nonumber \\
\underbar{\ensuremath{\mathbb{M}}}_{OI}\mathbb{X}_{OI}(T) & = & \mathbb{S}_{OI}(T).\label{eq:MX=00003DS}
\end{eqnarray}

Clearly, Eq. (\ref{eq:MX=00003DS}) can be written as a single matrix
equation \noun{$\mathbb{\underbar{\ensuremath{\mathbb{M}}}}\mathbb{\mathcal{\mathbb{X}}}(T)=\mathbb{S}(T)$,}\emph{
}where $\underbar{\ensuremath{\mathbb{M}}}=\underbar{\ensuremath{\mathbb{M}}}_{OO}\bigoplus\underbar{\ensuremath{\mathbb{M}}}_{II}\bigoplus\underbar{\ensuremath{\mathbb{M}}}_{OI}$
is a $24\times16$ matrix with each of the original matrices along
the diagonal and zeros for the rest of the entries, i.e., it is of
the block-diagonal form. $\mathbb{\mathcal{\mathbb{X}}}(T)$ and $\mathbb{S}(T)$
are the concatenations of the corresponding column vectors and have
sizes $16$ and $24$, respectively. The condition number of $\underbar{\ensuremath{\mathbb{M}}}$
is equal to 14.9, which indicates a well-behaved inversion, associated
with the sparsity of the matrix%
\footnote{The smaller the condition number, the more stable the inversion; since
some signals are directly proportional to certain elements of $\chi(T)$,
the matrix is sparse and the inversion is stable.%
}. Yet a naive direct inversion of $\underbar{\ensuremath{\mathbb{M}}}$
yields unphysical values of the process matrix $\chi(T)$ (in this
case, of the vector $\mathbb{\mathcal{\mathbb{X}}}(T)$) in general.
Via a semidefinite programming routine built using the CVX software
\cite{gb08,cvx}, we impose the positive-semidefinite constraint,

\begin{equation}
\sum_{ijqp}z_{iq}^{*}\chi_{ijqp}(T)z_{jp}\geq0,\label{eq:pos_s}
\end{equation}
for any complex-valued matrix $z$. This condition guarantees that
the inverted $\chi(T)$ maps positive density matrices to other positive
density matrices. The result of this numerical procedure is given
in Fig. 5 in main text, where most of the elements of $\chi(T)$ (namely,
the nonsecular terms) result to be negligible.

\item Since we do not precisely know the dipoles of the system, we now test
the sensitivity of the extracted $\chi(T)$ to the dipole coefficients
$\mathbb{\underbar{\ensuremath{\mathbb{M}}}}$ (Eq. (\ref{eq:matrix_M})).
We modify $\underbar{\ensuremath{\mathbb{M}}}$ by scaling one of
the coefficients by a factor and keeping the rest fixed. This generates
a matrix $\underbar{\ensuremath{\mathbb{M}}}'$ from which we can
extract a new $\mathbb{\mathcal{\mathbb{X}}}'(T)$. We compute two
error measures associated with this perturbation,

\begin{equation}
\mbox{Error}_{1}(\underbar{\ensuremath{\mathbb{M}}}')=\frac{\mbox{max}_{ijqp}\sum_{T}|\chi_{ijqp}(T)-\chi_{ijqp}'(T)|}{\mbox{Number of \ensuremath{T}points}},\label{eq:Error}
\end{equation}

\begin{equation}
\mbox{Error}_{2}(\underbar{\ensuremath{\mathbb{M}}}')=\frac{\mbox{max}_{T}\sum_{ijqp}|\chi_{ijqp}(T)-\chi_{ijqp}'(T)|}{\mbox{Number of elements in \textsc{\ensuremath{\mathbb{\mathcal{\mathbb{X}}}}(T)}}},\label{eq:Error-1}
\end{equation}

where the $\mbox{Number of \ensuremath{T}points}$ is 33 (from 0 to
510 fs) and the $\mbox{Number of elements in \textsc{\ensuremath{\mathbb{\mathcal{\mathbb{X}}}}(T)}}$
is 16 (the size of $\mathbb{\mathcal{\mathbb{X}}}(T)$). The results
of these calculations are shown in Table S2-1 and 2. Notice that $\mbox{Error}_{2}(\underbar{\ensuremath{\mathbb{M}}}')$
is in general smaller than $\mbox{Error}_{1}(\underbar{\ensuremath{\mathbb{M}}}')$.
This has to do with the fact that most of the elements of $\chi(T)$,
the nonsecular terms, are negligible, and varying the coefficients
of $\underbar{\ensuremath{\mathbb{M}}}$ keeps them that way, which
is a good sign. Since $\mbox{Error}_{2}(\underbar{\ensuremath{\mathbb{M}}}')$
averages the error over the different elements of $\chi(T)$, the
nonsecular terms ``buffer'' the errors from the other terms. On
the other hand, $\mbox{Error}_{1}(\underbar{\ensuremath{\mathbb{M}}}')$
averages instead over the $T$ points, and singles out the highest
deviation amongst the different elements of $\chi(T)$.

Table S2-1 shows that our extracted process matrices are most sensitive
to deviations in the coefficient $D$. In particular, scaling $D$
by a factor of -1.6, -4 or $\pm10$ can cause significant deviations
in at least one element of $\chi(T)$. Multiplying $D$ by a factor
of 1.6 does not cause large changes in $\chi(T)$. In evaluating $D$,
we assume that $\mu_{II,I}=\mu_{Ig}$. This approximation could be
checked by quantum chemistry calculations. The other dipole combinations
must be wrong by at least a factor of $\pm10$ before significant
errors in $\chi(T)$ are introduced. And the results seem largely
independent of the value of $F$ entirely. These observations, together
with the relatively small condition number of $\mathbb{M}$, indicate
that the inversion of $\mathbb{\mathcal{\mathbb{X}}}(T)$, and hence
the QPT, is not too sensitive to the precise values of the coefficients
of $\underbar{\ensuremath{\mathbb{M}}}$.

\begin{center}
\begin{tabular}{|l|c|c|c|c|c|c|c|c|c|c|c|c|c|}
\hline
\multicolumn{14}{|l}{TABLE S2-1. Sensitivity analysis of errors on the coefficients of
$\underbar{\ensuremath{\mathbb{M}}}$ ($\mbox{Error}{}_{1}$)}\tabularnewline
\hline
Coefficient\textbackslash{}Scaling factor & \textbf{-10} & \textbf{-4} & \textbf{-1.6} & \textbf{-0.6} & \textbf{-0.25} & \textbf{-0.1} & \textbf{0.1} & \textbf{0.25} & \textbf{0.6} & \textbf{1} & \textbf{1.6} & \textbf{4} & \textbf{10}\tabularnewline
\hline
$A$ & 0.38 & 0.39 & 0.39 & 0.36 & 0.21 & 0.13 & 0.06 & 0.09 & 0.06 & 0 & 0.12 & 0.28 & 0.34\tabularnewline
\hline
$B$ & 0.81 & 0.14 & 0.01 & 0.01 & 0.01 & 0.01 & 0 & 0 & 0 & 0 & 0.01 & 0.16 & 0.79\tabularnewline
\hline
$C$ & 0.37 & 0.15 & 0.07 & 0.04 & 0.03 & 0.02 & 0.02 & 0.01 & 0.01 & 0 & 0.01 & 0.09 & 0.32\tabularnewline
\hline
$D$ & 0.99 & 0.98 & 0.9 & 0.81 & 0.77 & 0.68 & 0.24 & 0.02 & 0.01 & 0 & 0.21 & 0.67 & 0.87\tabularnewline
\hline
$E$ & 0.21 & 0.23 & 0.03 & 0.05 & 0.07 & 0.07 & 0.07 & 0.07 & 0.06 & 0 & 0.03 & 0.15 & 0.18\tabularnewline
\hline
$F$ & 0.04 & 0.02 & 0.01 & 0.01 & 0.01 & 0.01 & 0.01 & 0.01 & 0.01 & 0 & 0.01 & 0.02 & 0.04\tabularnewline
\hline
$G$ & 0.22 & 0.26 & 0.37 & 0.06 & 0.06 & 0.06 & 0.05 & 0.04 & 0.02 & 0 & 0.05 & 0.12 & 0.16\tabularnewline
\hline
$H$ & 0.21 & 0.23 & 0.05 & 0.03 & 0.07 & 0.07 & 0.07 & 0.07 & 0.05 & 0 & 0.03 & 0.15 & 0.18\tabularnewline
\hline
\end{tabular}
\par\end{center}

\begin{center}
\begin{tabular}{|l|c|c|c|c|c|c|c|c|c|c|c|c|c|}
\hline
\multicolumn{14}{|l}{TABLE S2-2. Sensitivity analysis of errors on the coefficients of
$\underbar{\ensuremath{\mathbb{M}}}$ ($\mbox{Error}{}_{2}$)}\tabularnewline
\hline
Coefficient\textbackslash{}Scaling factor & \textbf{-10} & \textbf{-4} & \textbf{-1.6} & \textbf{-0.6} & \textbf{-0.25} & \textbf{-0.1} & \textbf{0.1} & \textbf{0.25} & \textbf{0.6} & \textbf{1} & \textbf{1.6} & \textbf{4} & \textbf{10}\tabularnewline
\hline
$A$ & 0.2 & 0.2 & 0.19 & 0.17 & 0.1 & 0.1 & 0.06 & 0.06 & 0.04 & 0 & 0.05 & 0.13 & 0.17\tabularnewline
\hline
$B$ & 0.13 & 0.04 & 0.01 & 0.01 & 0.01 & 0.01 & 0.01 & 0.01 & 0 & 0 & 0 & 0.03 & 0.08\tabularnewline
\hline
$C$ & 0.19 & 0.07 & 0.03 & 0.01 & 0.01 & 0.01 & 0.01 & 0.01 & 0 & 0 & 0.01 & 0.05 & 0.14\tabularnewline
\hline
$D$ & 0.14 & 0.13 & 0.13 & 0.13 & 0.13 & 0.07 & 0.04 & 0.01 & 0.01 & 0 & 0.04 & 0.09 & 0.11\tabularnewline
\hline
$E$ & 0.09 & 0.11 & 0.02 & 0.03 & 0.04 & 0.05 & 0.05 & 0.05 & 0.03 & 0 & 0.02 & 0.07 & 0.08\tabularnewline
\hline
$F$ & 0.02 & 0.01 & 0.01 & 0 & 0.01 & 0.01 & 0.01 & 0.01 & 0 & 0 & 0.01 & 0.01 & 0.02\tabularnewline
\hline
$G$ & 0.1 & 0.11 & 0.16 & 0.04 & 0.03 & 0.03 & 0.03 & 0.02 & 0.02 & 0 & 0.03 & 0.06 & 0.07\tabularnewline
\hline
$H$ & 0.1 & 0.11 & 0.03 & 0.02 & 0.04 & 0.04 & 0.05 & 0.05 & 0.03 & 0 & 0.02 & 0.08 & 0.08\tabularnewline
\hline
\end{tabular}
\par\end{center}

\end{enumerate}
\bibliographystyle{unsrt}